\documentclass[prd,aps,showpacs,eqsecnum,preprint]{revtex4}
\usepackage{graphics,epsfig}
\begin{document}

\count255=\time\divide\count255 by 60 \xdef\hourmin{\number\count255}
  \multiply\count255 by-60\advance\count255 by\time
 \xdef\hourmin{\hourmin:\ifnum\count255<10 0\fi\the\count255}

\newcommand{\xbf}[1]{\mbox{\boldmath $ #1 $}}

\newcommand{\sixj}[6]{\mbox{$\left\{ \begin{array}{ccc} {#1} & {#2} &
{#3} \\ {#4} & {#5} & {#6} \end{array} \right\}$}}

\newcommand{\threej}[6]{\mbox{$\left( \begin{array}{ccc} {#1} & {#2} &
{#3} \\ {#4} & {#5} & {#6} \end{array} \right)$}}

\title{Constraints on Natural MNS Parameters from $|U_{e3}|$}

\author{Richard F. Lebed}
\email{Richard.Lebed@asu.edu}

\author{Daniel R. Martin}
\email{Daniel.Martin@asu.edu}

\affiliation{Department of Physics and Astronomy, Arizona State
University, Tempe, AZ 85287-1504}

\date{December, 2003}

\begin{abstract}
The MNS matrix structure emerging as a result of recent neutrino
measurements strongly suggests two large mixing angles (solar and
atmospheric) and one small angle ($|U_{e3}| \! \ll \! 1$).  Especially
when combined with the neutrino mass hierarchy, these values turn out
to impose rather stringent constraints on possible flavor models
connecting the three active fermion generations.  Specifically, we
show that an extremely small value of $|U_{e3}|$ would require fine
tuning of Majorana mass matrix parameters, particularly in the context
of seesaw models.
\end{abstract}


\maketitle

\section{Introduction}

Recent progress in experimental neutrino physics has been nothing
short of breathtaking.  At the time of this writing, only five years
have elapsed since Super-Kamiokande's observation~\cite{SKatm} of the
atmospheric neutrino deficit.  Just last year the SNO experiment
showed~\cite{SNO} through detection of neutral current events that the
solar neutrino deficit, as observed by decades of steadily improving
charged current experiments (Homestake~\cite{Homestake},
Kamiokande~\cite{Kamiokande}, SAGE~\cite{SAGE}, GALLEX~\cite{GALLEX},
GNO~\cite{GNO}, Super-Kamiokande~\cite{SKsolar}, and SNO
itself~\cite{SNO}) is truly a deficit due to solar $\nu_e$'s
oscillating into other active flavors.  In the past year the K2K
experiment~\cite{K2K} confirmed nicely the atmospheric mixing
parameters measured by Ref.~\cite{SKatm} using long-baseline
accelerator techniques, while the KamLAND reactor
experiment~\cite{KamLAND} tightly constrained the mixing parameter
space previously probed only through solar neutrino measurements, and
the WMAP satellite experiment~\cite{WMAP} used cosmological
constraints to bound the neutrino mass sum to lie below $0.71$~eV (at
95\% CL).

This list does not do justice to the large number of completed
neutrino experiments that provide important exclusionary bounds, nor
the panoply of new experiments underway or in the planning stages.
Our intent is merely to provide a snapshot of a rapidly maturing
field, and to point out that, even now, we possess a reasonably good
picture of neutrino masses and mixing as encapsulated by the
Maki-Nakagawa-Sakata (MNS) matrix.  According to a recent three-flavor
analysis~\cite{MSTV}, the parameters with 2$\sigma$ uncertainties are
given by
\begin{eqnarray}
| \Delta m_{32}^2 | & = & (1.8-3.3) \times 10^{-3} {\rm eV}^2 ,
\label{Deltam32} \\
\sin^2 \theta_{23} & = & 0.36-0.67 , \label{atmmix} \\
\Delta m_{21}^2 & = & (6.0-8.4) \times 10^{-5} {\rm eV}^2 ,
\label{Deltam21} \\
\sin^2 \theta_{12} & = & 0.25-0.36 , \label{solmix} \\
\sin^2 \theta_{13} & \le & 0.035 , \label{theta13}
\end{eqnarray}
where the magnitude symbols on $\Delta m_{32}^2$ serve as a reminder
that either the normal or inverted hierarchy is still allowed.  The
important point here is not the precise values of the observables
(which seem to improve monthly in response to new data regularly being
released), but rather the picture painted by their magnitudes: All
neutrino masses are small, their mass differences are smaller still
and hierarchical, and precisely two of the three Euler angles in the
MNS matrix are large (``bimaximal''), while the third is quite small
(a result of including the CHOOZ~\cite{CHOOZ} and Palo Verde~\cite{PV}
exclusionary data in the global analysis).  This structure differs
radically from that found in the quark sector, where the CKM matrix
possesses three small and hierarchically decreasing Euler angles ({\it
i.e.}, $1 \! \gg \! |V_{us}| \! \gg \!  |V_{cb}| \! \gg \! |V_{ub}|$).

It is natural to suppose that the existence of three complete flavors
(generations) of spin-$\frac 1 2$ fermions, with members possessing
the same quantum numbers under all standard model gauge groups and
differing only in mass, suggests some sort of flavor symmetry.  This
is, after all, how the periodic table of the elements and the quark
structure of hadrons came to be understood.  In the case of an
ostensible flavor symmetry, however, one must incorporate a number of
disparate properties: Masses extend from $m_t \simeq 174$~GeV, near
the weak scale, through those of a number of light quarks and charged
leptons in the MeV to GeV range, down to the neutrinos, with masses
(perhaps) in the meV range.  A flavor symmetry must also accommodate
the aforementioned differences between the CKM and MNS matrices.
Finally, if the flavor symmetry leaves a signature detectable at weak
scales (such as in the context of supersymmetry), it must avoid
creating unacceptably large flavor-changing neutral currents.

In the case of the neutrinos, the seesaw mechanism provides a natural
machinery for obtaining tiny mass eigenvalues, and in the following we
employ the mechanism in its simplest form.  Of course, the seesaw also
requires neutrinos with a Majorana character, and the presence of
Majorana in addition to Yukawa (Dirac) mass matrices both complicates
and enriches the possible theory parameter space.  For our current
purposes, however, we are interested foremost in the general nature of
the (derived) left-handed Majorana mass matrix as well as those of the
Yukawa and right-handed Majorana matrices that generate it, but less
so in their fundamental origins.  Although the construction of a model
to describe the origin of neutrino couplings is theoretically
important, one must consider the low-energy phenomenological
consequences of the model first in order to understand its
limitations.  The emerging precision of findings such as those in
Eqs.~(\ref{Deltam32})--(\ref{theta13}), based on data only a few
months old, suggests that we are rapidly approaching a ``tipping
point'' at which exceptionally few Yukawa and Majorana matrix
textures, and consequently a rarified set of possible flavor
symmetries, will remain viable.

The purpose of this paper is to explore constraints on natural
textures that simultaneously give rise to bimaximal atmospheric
($\theta_{23}$) and solar ($\theta_{12}$) angles, while maintaining a
small $\theta_{13}$ and a mass splitting hierarchy $\Delta
m_{21}^2/|\Delta m_{32}^2| \ll 1$.  We show, using only naturalness
criteria such as the assumption that the small value of $\theta_{13}$
is not due to a fortuitous cancellation between charged lepton Yukawa
and neutrino Majorana parameters, that there is an effective lower
bound on the natural size of $\theta_{13}$ values arising from flavor
models that predict the elements of the Majorana and Yukawa matrices.
In particular, $\theta_{13}$ cannot be much smaller than its currently
measured bound, unless one resorts to fine tuning in either or both of
the Yukawa and Majorana matrices.  A recent paper~\cite{IR}, in the
same spirit of studying mass matrix textures, also finds that
$\theta_{13}$ should lie close to its current bound; however,
Ref.~\cite{IR} places emphasis on texture zeroes instead of (as here)
the naturalness constraints of allowing matrix elements to be as large
as possible.

We note that the prediction that $\theta_{13}$ is not expected to be
excessively small has been made in several previous
works~\cite{ABR,FSV,LMS,KingUe3}, in the context of various models and
parametrizations.  Moreover, Ref.~\cite{BD} studies this issue by
examining classes of models appearing in the literature and finds a
similar effect.  In this paper we derive and explain a simple
underlying reason for this recurring result.

The paper is organized as follows: In Sec.~\ref{conv} we establish the
conventions defining the relevant matrices.  In Sec.~\ref{conts} we
lay out the naturalness criteria used to study Yukawa and Majorana
textures and see how they are realized in a sample model.
Section~\ref{build} presents a constructive method of building the
left-handed neutrino Majorana matrix that satisfies all existing data,
at which point we observe how little effect a sufficiently small
$\theta_{13}$ would have on the textures.  Sections~\ref{NH} and
\ref{IH} explore the consequences of these constraints for the
textures of the heavy right-handed neutrinos entering the seesaw
mechanism in the normal and inverted hierarchies, respectively.
Section~\ref{concl} concludes.

\section{Conventions and Notations} \label{conv}

The conventions defining the charged lepton Yukawa matrix $Y_L$, the
neutrino Yukawa matrix $M_{LR}$, and the neutrino Majorana mass
matrices $M_{LL}$ and $M_{RR}$ are established simply by presenting
the relevant Lagrangian terms.  What follows is of course well known,
but is included in order to leave no ambiguities in the notation used
in the rest of the paper.  Beginning with the neutrino terms,
\begin{eqnarray}
{\cal L}_\nu & = &
\bar \nu_L i \! \! \not \! \partial \nu_L \! +
\bar \nu_R i \! \! \not \! \partial \nu_R
- \! \frac 1 2 \left( \bar \nu^c_L M_{LL} \nu_L + {\rm H.c.} \right)
- \! \frac 1 2 \left( \bar \nu^c_R M_{RR} \nu_R + {\rm H.c.} \right)
- \left( \bar \nu_L M_{LR} \nu_R + {\rm H.c.} \right) , \nonumber \\
& &
\end{eqnarray}
where $\psi^c \! \equiv C \bar \psi^T$ is the field obtained via the
charge conjugation operation $C$, H.c.\ = Hermitian conjugate, and the
fields $\nu_{L,R}$ stand for spinors containing all three neutrino
generations, one first notes that the presence of a transposed (rather
than Hermitian-conjugated) field in the ket of Majorana terms
constrains $M_{LL}$ and $M_{RR}$ both to be (complex) symmetric
matrices.  In the most minimal seesaw mechanism, $M_{LL}$ does not
enter as a fundamental field in the Lagrangian, but rather arises as
an effective operator:
\begin{equation}
M^{\rm eff}_{LL} = -M_{LR}^{\ast} M_{RR}^{-1} M_{LR}^\dagger \ .
\label{seesaw}
\end{equation}
As discussed in the Introduction, in this work we begin by treating
the effective $M_{LL}$ precisely as one would any fundamental
Lagrangian parameter.

The Lagrangian in the lepton sector is then defined by:
\begin{eqnarray}
{\cal L} & = &
\bar \nu_L i \! \! \not \! \partial \nu_L \! +
\bar e_L i \! \! \not \! \partial e_L +
\bar e_R i \! \! \not \! \partial e_R
- \frac 1 2 \left( \bar \nu^c_L M_{LL} \nu_L + {\rm H.c.} \right)
- \frac{v}{\sqrt{2}} \left( \bar e_L Y_L e_R + {\rm H.c.} \right)
\nonumber \\ & & + \frac{g_W}{\sqrt{2}}
\left( \bar e_L \! \not \! \! {\cal W}^- \nu_L + {\rm H.c.} \right) \ ,
\end{eqnarray}
where $g_W$ is the usual electroweak gauge coupling and $v \! \simeq
\!  246$~GeV.  We diagonalize $Y_L$ in the usual way, with a biunitary
transformation given by
\begin{equation}
Y_L^0 \! = \! U_L^\dagger Y_L U_R \ ,
\label{CLT} 
\end{equation} 
where $Y_L^0$ is diagonal.  That is, the mass basis is related to the
weak interaction basis by $e_L^0 \! = \! U_L^\dagger e_L$, $e_R^0 \! =
\! U_R^\dagger e_R$, and $U_L$ is obtained by diagonalizing
$Y_L Y_L^\dagger$:
\begin{eqnarray}
Y_L^0 Y_L^{0 \dagger} & = & U_L^\dagger Y_L U_R \, U_R^\dagger
Y_L^\dagger U_L \nonumber \\
& = & U_L^\dagger Y_L Y_L^\dagger U_L \ ,
\label{CLT2}  
\end{eqnarray}
while $U_R$ may be obtained by diagonalizing $Y_L^\dagger Y_L$.

Since the most general Majorana mass matrix $M_{LL}$ is complex
symmetric rather than Hermitian, the usual procedure of
diagonalization via a unitary matrix composed of its eigenvectors is
no longer entirely valid.  Nevertheless, one can prove~\cite{norm}
that for any such $M_{LL}$ there exists a unitary matrix $W$ such that
\begin{equation}
M_{LL}^0 \! = \! W^T M_{LL} W
\label{LHNM}
\end{equation}
is diagonal (in mathematical jargon, ``is reduced to normal form''),
and $\nu_L^0 \! = \! W^\dagger \nu_L$.  The kinetic and mass terms are
unaffected by these transformations, and the interaction term becomes
\begin{equation}
{\cal L}_{\rm int} = + \frac{g_W}{\sqrt{2}}
\left( \bar e_L^0 \! \not \! \! {\cal W}^- U_L^\dagger W \nu_L^0
+ {\rm H.c.} \right) , \label{WeakMass}
\end{equation}
from which one defines the MNS matrix as
\begin{equation} 
U_{\rm MNS} = U_L^\dagger W \ . 
\label{MNS}
\end{equation}

The full MNS matrix provides the linear combination of freely
propagating neutrino mass eigenstates that interacts at a weak vertex
with a particular charged lepton mass eigenstate.  For example,
Eqs.~(\ref{WeakMass}) and (\ref{MNS}) show that an electron is
produced from the ``isolated'' neutrino mass eigenstate $\nu_3^0$ with
an amplitude given by (dropping the MNS subscript) the now-famous
matrix element $U_{e3}$.

Although the construction of $U_{\rm MNS}$ closely resembles that of
the CKM matrix, the Majorana nature of the neutrinos permits two
additional observable phases.  Using abbreviations $\sin \theta_{ij}
\! \equiv \!  s_{ij}$ and $\cos \theta_{ij} \! \equiv \! c_{ij}$, one
convenient parametrization~\cite{King} for $U_{\rm MNS}$ reads
\begin{equation}
U_{\rm MNS} = \left( \begin{array}{ccc}
c_{13} c_{12} \, e^{i\beta_1} &
c_{13} s_{12} \, e^{i\beta_2} &
s_{13} \, e^{-i\delta} \\
(-s_{12} c_{23} - s_{13} c_{12} s_{23} \, e^{i\delta}) \, e^{i\beta_1} &
(+c_{12} c_{23} - s_{13} s_{12} s_{23} \, e^{i\delta}) \, e^{i\beta_2} &
c_{13} s_{23} \\
(+s_{12} s_{23} - s_{13} c_{12} c_{23} \, e^{i\delta}) \, e^{i\beta_1} &
(-c_{12} s_{23} - s_{13} s_{12} c_{23} \, e^{i\delta}) \, e^{i\beta_2} &
c_{13} c_{23}
\end{array} \right) . \label{MNSparam}
\end{equation}
The two CP-violating parameters $\beta_{1,2}$ represent relative
phases between the neutrino mass eigenstates and can be removed from
$U_{\rm MNS}$ in Eqs.~(\ref{WeakMass})--(\ref{MNS}); they do not
contribute to neutrino oscillation measurements~\cite{BHP}, leaving
$\delta$ as the sole CP-violating parameter discernable in such data.
However, the Majorana mass term is not invariant under such a
redefinition of phases, meaning that only experiments sensitive to the
Majorana nature of neutrinos, such as neutrinoless double beta decay,
can probe $\beta_{1,2}$.  Notably, the conventions are chosen so that
the phase $\delta$ multiplies the smallest Euler angle $\theta_{13}$,
thus minimizing its impact in the texture.

\section{Contributions to the MNS matrix} \label{conts}

\subsection{General Features}

As is clear from Eq.~(\ref{MNS}), the MNS matrix receives
contributions from both the unitary matrix $U_L$ that diagonalizes the
charged lepton Yukawa matrix combination $Y_L Y_L^\dagger$ and the
unitary matrix $W$ that puts the neutrino Majorana mass matrix
$M_{LL}$ into normal form.  In general, both matrices can contribute
equally to the observed mixing angles.  However, as we argue in this
section, such a democratic structure leads to either phenomenological
or theoretical problems if one supposes that a natural flavor symmetry
underlies these textures.  To be specific, it does not appear possible
to satisfy simultaneously the following criteria:
\begin{enumerate}

\item The Yukawa matrices of all light fermions---at least the
charged ones---share a common origin in some flavor symmetry.

\item Hierarchies of charged fermion masses and quark mixing angles
have a common origin, arising through the presence of small flavor
symmetry-breaking parameters occupying certain elements of the Yukawa
matrices.

\item There are no excessive fine-tuning cancellations between
elements of $U_L$ and $W$ in forming $U_{\rm MNS}$ (or the unitary
matrices that diagonalize the two left-handed quark Yukawa matrices
and that comprise the CKM matrix).

\item The value of $|U_{e3}|$ is very small (in a sense made explicit
below).

\end{enumerate}

Point~(1) is the assertion that the texture of the Yukawa matrix $Y_L$
is the same as those of $Y_U$ and $Y_D$ in the quark sector:
Corresponding elements of $Y_L$, $Y_U$, and $Y_D$ differ numerically,
of course, but have the same order of magnitude.  Such a pattern
occurs very naturally in grand unified theories.  If this assumption
fails, then the basic assumption that a flavor symmetry connects all
fermions in a given generation falls into doubt.  Point~(2) asserts
that the observed hierarchies, such as $m_e \! \ll \! m_{\mu} \! \ll
\! m_{\tau}$ and $1 \! \gg \! |V_{us}| \! \gg \! |V_{cb}| \! \gg \!
|V_{ub}|$, arise from powers of the {\em same\/} small parameters
(which are as few in number as possible); this of course is the usual
miminality assumption.  Point~(3) is a naturalness assumption, that no
small numbers should arise unless placed into the theory by hand.

Now, the fact that the each CKM angle is small, or equivalently that
the CKM matrix is only perturbatively different from the identity
matrix $\openone$, implies that [by Point~(3)] the separate unitary
matrices that diagonalize the quark Yukawa matrices $Y_U Y_U^\dagger$
and $Y_D Y_D^\dagger$ are each only perturbatively different from
being diagonal in the same basis.  This, in turn, implies that $Y_U
Y_U^\dagger$ and $Y_D Y_D^\dagger$ are each only perturbatively
different from being diagonal themselves~\footnote{This is true in the
basis in which $Y_U Y_U^\dagger$ and $Y_D Y_D^\dagger$ are both
diagonal.  If one considers only the low-energy theory, it is always
possible to rotate into this basis; however, for a given high-energy
theory this freedom is lost since it may have its own preferred basis
for Yukawa couplings, and the unitary matrix connecting the two bases
can contain interesting information.  See, {\it e.g.},
Ref.~\cite{Derm}.}.  By the same reasoning, it would also be unnatural
for a Yukawa combination such as $Y_U Y_U^\dagger$ to be nearly
diagonal and yet for $Y_U$ itself to have large off-diagonal elements.

The notion that the quark Yukawa matrices themselves differ only
perturbatively from being diagonal matrices is a feature common to
virtually all flavor models.  Indeed, perhaps the prettiest prediction
of flavor models (and its oldest~\cite{VusPred}) is the relation
$V_{us} \! \approx \! \sqrt{m_d/m_s}$, which is obtained by supposing
that the upper $2 \! \times \! 2$ block of $Y_D$ assumes the basic
form
\begin{equation}
\left. Y_D^{\vphantom{\dagger}} \right|_{2\times2} =
\left( \begin{array}{cc} 0 & O(\alpha) \\ O(\alpha) & 1
\end{array} \right) , \label{12entry}
\end{equation}
where $\alpha \! = \! O(V_{us}) \! = \! O(0.1)$ is a small flavor
symmetry-breaking parameter, leading to a ratio of $m_d$ to $m_s$
eigenvalues in the phenomenologically correct proportion $\alpha^2 \!
: \! 1$.  The relation of CKM and quark mass parameters is, of course,
an example of Point~(2).

Point~(1) then suggests that the same basic texture applies to $Y_L$:
No large mixing angles should arise in $Y_L$.  The observed $O(1)$
atmospheric and solar mixing angles [Eqs.~(\ref{atmmix}) and
(\ref{solmix}), respectively], combined with Eq.~(\ref{MNS}), implies
that all large mixing angles must arise solely from $W$.  While it is
then tempting to suppose that the texture of $U_{\rm MNS}$ is
determined entirely by $W$, there is still the matter of the small
element $U_{e3}$, to which we next turn.

In order to distinguish contributions to $U_{e3}$ from the charged
lepton and neutrino sectors, one may adopt a separate parametrization
for each of $U_L$ and $W$ (for which each angle is labeled with an
$E$ or $\nu$ superscript, respectively) analogous to that of the full
$U_{\rm MNS}$ in Eq.~(\ref{MNSparam}).  Then, up to first order in the
small angles $\theta_{12}^E$, $\theta_{13}^E$, $\theta_{23}^E$, and
$\theta_{13}^\nu$, a straightforward calculation gives
\begin{equation} \label{Ue3exp}
U_{e3} = s_{13} \, e^{-i\delta} \simeq
[\theta_{13}^\nu \, e^{-i\delta^\nu} \!\!
- \theta_{13}^E \, c_{23}^\nu \, e^{-i\delta^E} \!\!
- \theta_{12}^E s_{23}^\nu ] \, e^{-i\beta_1^E} .
\end{equation}
An analogous expression using different phase conventions appears in
Ref.~\cite{King}.  Regardless of the specific form, however, the
important point is that $|U_{e3}|$ is generically as large as the
largest of $\theta_{13}^\nu$, $\theta_{13}^E$, and $\theta_{12}^E$.
The phenomenological fact from Eq.~(\ref{theta13}) that $|U_{e3}| \!
\alt \! O(0.1)$ provides additional evidence that at least two of the
charged lepton mixing angles, $\theta_{13}^E$ and $\theta_{12}^E$, are
small.  Even so, from this reasoning we see that $U_{e3}$ may receive
important contributions from $Y_L$.  If one takes Point~(1) at face
value, and supposes that the reasoning leading to Eq.~(\ref{12entry})
applies to $Y_L$, then Point~(3) demands that $|U_{e3}|$ must be at
least $O(\alpha)$, since the $O(\alpha)$ element $(Y_L)_{21}$ induces
through the diagonalization process an $O(\alpha)$ element
$(U_L)_{21}$, which implies that $\theta_{12}^E \! = \!  O(\alpha)$,
and thus by Eq.~(\ref{Ue3exp}), $|U_{e3}|$ is generically at least
$O(\alpha)$.  One concludes that the current upper bound
Eq.~(\ref{theta13}) for $\sin^2 \theta_{13}$, which is about
$O(\alpha^2)$, cannot greatly exceed its actual value, for then
Point~(4) would be violated.

One might now argue that the assumptions of Points~(1)--(3) are
perhaps too restrictive; after all, perhaps there is a mild tuning of
the $O(1)$ coefficients in the Yukawa matrices that leads to a
somewhat smaller $(U_L)_{21}$ element, and hence a smaller
$\theta_{12}^E$.  Or, perhaps there exist other natural Yukawa
textures in the quark sector that do not use the mechanism of
Eq.~(\ref{12entry}).  Indeed, for the remainder of this paper, let us
suppose that there exist some means by which one can render the
contributions of $Y_L$ to $U_{e3}$ at most of the order of
contributions from $W$.  In that case then, at least in terms of
orders of magnitude, we are making the assumption
\begin{equation} \label{MNSapprox}
U_{\rm MNS} \simeq W \ .
\end{equation}
Our goal in the remainder of this paper is to show that, even in this
scenario, obtaining a value for for $U_{e3} \simeq W_{13}$ as small as
$O(\alpha^2)$ is unnatural.  Specifically, for a model that predicts
elements of $M_{LL}$ (or equivalently $M_{LR}$ and $M_{RR}$) to
produce $U_{e3} \! = \! O(\alpha^2)$ requires a degree of fine tuning
beyond the leading $\alpha$ orders of the matrix elements.

\subsection{Specific Realization}

In order to see how this plays out in a full-fledged flavor model, we
first introduce a bit of notation useful in discussing the problem.
Let us begin by introducing a single universal small parameter $\rho
\! < \! 1$ to describe flavor symmetry breaking.  In typical models, the
assumed flavor group $G_{\! f}$ with ultraviolet cutoff $M_f$ exhibits
a sequential symmetry-breaking pattern to subgroups $G_{\! f} \!
\supset \! H_1 \! \supset \! H_2 \! \supset \! \cdots$,
\begin{equation}
G_{\! f} \stackrel{M_{\! f} \rho^{m_1}}{\longrightarrow} H_1
\stackrel{M_{\! f} \rho^{m_2}}{\longrightarrow} H_2
\stackrel{M_{\! f} \rho^{m_3}}{\longrightarrow} \cdots \, ,
\end{equation}
where the integers $0 \! < \! m_1 \! < \! m_2 \! < \! \cdots$ label
sequential scales of symmetry breaking, $M_f \! > M_f \rho^{m_1} \!  >
\cdots$.  Models with multiple scales can then be represented by
identifying the integers $m_i$ most closely associated with the order
of magnitude appropriate to each step of symmetry breaking.

Models based on $G_f$ = U(2)~\cite{U2} and its discrete subgroup
$T^\prime$~\cite{ACL,ACM} exhibit two steps of symmetry breaking, with
associated dimensionless parameters $\epsilon \! \approx \! 0.04$ and
$\epsilon^\prime \! \approx \! 0.004$.  Equivalently, one may take
$\rho \approx 0.2$ and $m_1 \! = \! 2$, $m_2 \! = \! 3$: $\epsilon \!
= \! O(\rho^2)$ and $\epsilon^\prime \! = \! O(\rho^3)$.  The charged
lepton Yukawa texture for these models is
\begin{eqnarray} 
Y_L \! \simeq \!
\left ( 
\begin{array}{ccc}
0                    & c_1 \epsilon^\prime & 0 \\
-c_1 \epsilon^\prime & 3 c_2 \epsilon      & c_3 \epsilon \\
0                    & c_4 \epsilon        & c_5 
\end{array} 
\right ) \ .
\label{ACL1}
\end{eqnarray}
The unitary matrix $U_L$ that diagonalizes $Y_L Y_L^\dagger$
[Eq.~(\ref{CLT2})] is given by
\begin{eqnarray} 
U_L \! \approx \! 
\left ( \begin{array}{ccc}
1  & \frac{c_1}{3c_2} \frac{\epsilon^\prime}{\epsilon} &
\frac{c_1 c_4}{c_5^2} \epsilon^\prime \epsilon \\
-\frac{c_1}{3c_2} \frac{\epsilon^\prime}{\epsilon} & 1 &
\frac{c_3}{c_5}  \epsilon \\
\frac{c_1 c_3}{3c_2 c_5} \epsilon^\prime  & -\frac{c_3}{c_5} \epsilon
& 1 
\end{array} \right ) ,
\label{LM1}
\end{eqnarray}
where only the leading-order contribution is retained in each element.
The charged leptonic mass texture is given by
\begin{eqnarray} 
Y_L^0 \! \approx \! 
\left ( 
\begin{array}{ccc}
\frac{c_1^2}{3c_2} \frac{\epsilon^{\prime \, 2}}{\epsilon} & 0 & 0 \\
0 & 3c_2\epsilon & 0 \\
0 & 0 &   c_5
\end{array} 
\right ) ,
\label{ChLeMa}
\end{eqnarray}
which agrees well with the observed hierarchy $m_e \! \ll \! m_\mu \!
\ll \!  m_\tau$.  However, from Eqs.~(\ref{MNS}), (\ref{MNSparam}), and
(\ref{LM1}), the element crucial to $U_{e3}$, namely $(U_L)_{21}$, is
$O(\epsilon^\prime/\epsilon) \! = \! O(\rho)$.  Thus, $\rho$ and the
parameter $\alpha$ previously used are of the same order of magnitude
(henceforth we use only $\rho$), and the problem of a large
contribution from $Y_L$ to $U_{e3}$ is manifested.  For example, in
the model of Ref.~\cite{ACM}, which was constructed to respect
bimaximal neutrino mixing, the most frequently obtained values of
$\theta_{13}$ cluster around 0.1~radians.

\section{Constructing $M_{LL}$} \label{build}

\begin{figure}[ht]


\epsfxsize 2.5 in \epsfbox{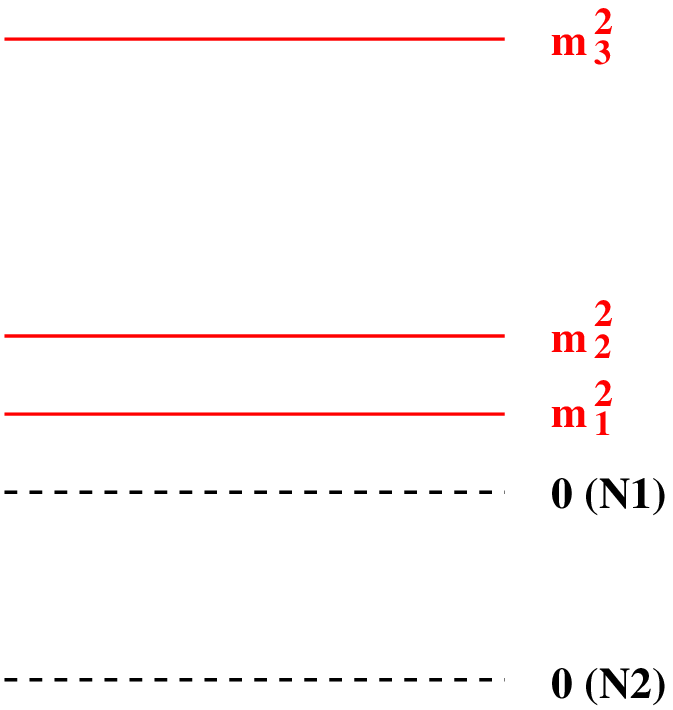} \hspace{5em}
\epsfxsize 2.5 in \epsfbox{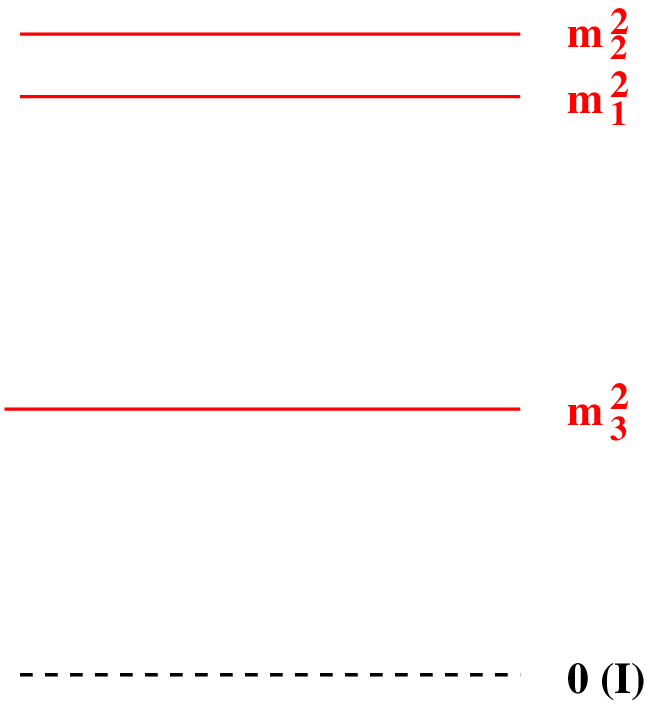}

\caption{
The normal ({\bf N1} and {\bf N2}) and inverted ({\bf I}) hierarchies.
Note that {\bf I} allows both large [$O(\rho^0)$] and small
[$O(\rho^1)$] $m_3$ values.}
\label{fig1}

\end{figure}

Under the ansatz of Eq.~(\ref{MNSapprox}), one can work backwards to
construct the texture of $M_{LL}$ via Eq.~(\ref{LHNM}):
\begin{equation} \label{back}
M_{LL} \simeq U_{\rm MNS}^{*} M_{LL}^0 U_{\rm MNS}^{\dagger} ,
\end{equation}
a calculation much in the spirit of Ref.~\cite{BHSSW}.  If one assumes
the seesaw mechanism Eq.~(\ref{seesaw}), then the absolute scale of
neutrino masses is determined in part by high-energy scale of
$M_{RR}$; in this work, however, we are interested only in ratios of
neutrino masses.  Using Eqs.~(\ref{Deltam32}) and (\ref{Deltam21}),
the key neutrino mass observable to accommodate is therefore
\begin{equation} \label{masshier}
r \equiv \Delta m_{21}^2/|\Delta m_{32}^2 | = 0.018-0.047 = O(\rho^2)
\ .
\end{equation}
This ratio has two well-known realizations, the normal ($m_1 \! < \!
m_2 \! < \! m_3$) and inverted ($m_3 \! < \! m_1 \! < \! m_2$)
hierarchies.  One must also ask how far the lightest neutrino mass
$m_0$ lies above zero, {\it i.e.}, whether it vanishes as some power
of $\rho$.  Factoring out the largest neutrino mass as an overall
scale, the possible normal hierarchy solutions to Eq.~(\ref{masshier})
are ({\bf N1}) $m_1 \! \equiv m_0 \! = O(\rho)$ or smaller, $m_2 \!
= \!  O(\rho)$, $m_3 \! = \!  O(\rho^0)$, and ({\bf N2}) $m_1 \!
\equiv m_0 \! = \! O(\rho^0)$, $m_2 \! = \!  m_0 \! + O(\rho^2)$,
$m_3 \! > \!  m_0$, while the inverted hierarchy solutions have ({\bf
I}) $m_1 \! \equiv m_0 \! = \!  O(\rho^0)$, $m_2 \! = \! m_0 \! +
O(\rho^2)$, and $m_3 \! = O(\rho^0)$ or smaller.  These schemes are
illustrated in Fig.~\ref{fig1}.


The key question then becomes this: Is it natural for a model to
produce a Majorana matrix $M_{LL}$ in any particular neutrino mass
hierarchy with $O(1)$ mixing angles $\theta_{12}$ and $\theta_{23}$,
but $\theta_{13} \! = O(\rho^2)$?  In this section we construct
$M_{LL}$ textures for each case.
\begin{eqnarray}
{\bf N1}: M_{LL}^0 & = & {\rm diag} \{ m_1^{(1)} \! \rho \! + \!
m_1^{(2)} \! \rho^2 , \; m_2^{(1)} \! \rho \! + \! m_2^{(2)} \! \rho^2
, \; m_3^{(0)}
\} \Rightarrow \nonumber \\  & & M_{LL} \sim \left(
\begin{array}{ccc} \rho & \rho & \rho \\ \rho & 1 & 1 \\ \rho & 1 & 1
\end{array} \right) \ , \label{N1}
\end{eqnarray}
(corresponding to Type IA in Ref.~\cite{King}), while
\begin{eqnarray}
{\bf N2}: M_{LL}^0 & = & {\rm diag} \{ m_0, \; m_0 \! \! +
\! m_2^{(2)} \! \rho^2, \; m_3 \} \ {\rm or} \nonumber \\
{\bf  I}: M_{LL}^0 & = & {\rm diag} \{ m_0, \; m_0 \! \! +
\! m_2^{(2)} \! \rho^2, \; m_3 \} \Rightarrow \nonumber \\
& & M_{LL} \sim \left( \begin{array}{ccc}
1 & 1 & 1 \\ 1 & 1 & 1 \\ 1 & 1 & 1 \end{array}
\right) \ . \label{inv}
\end{eqnarray}
Note that {\bf N2} and {\bf I} differ through the sign of $(m_3 \! -
\!  m_1$), but otherwise give the same texture for $M_{LL}$.  However,
while the given $M_{LL}$ textures seem simple and easily realizable
[particularly the seemingly anarchic structure of Eq.~(\ref{inv})],
they contain substantial internal correlations~\footnote{Possible
subleading contributions to $m_3$ in Eq.~(\ref{N1}) are removed by
redefining the overall (undetermined) scale of the matrix $M_{LL}$.
In the case of Eq.~(\ref{inv}) this cannot be done simultaneously for
$m_0$ and $m_3$, but subleading corrections to $m_3$, if desired, may
be inserted into the following expressions by hand.}.

To be specific, let $s_{13} \! = \! k \rho^2$.  Then, for {\bf N1},
the distinct elements read
\begin{eqnarray}
(M_{LL})_{11} & = & (m_1^{(1)} c_{12}^2 \, e^{-2i\beta_1} + m_2^{(1)}
s_{12}^2 \, e^{-2i\beta_2}) \rho + (m_1^{(2)} c_{12}^2 \,
e^{-2i\beta_1} + m_2^{(2)} s_{12}^2 \, e^{-2i\beta_2}) \rho^2 +
O(\rho^4) , \nonumber \\
(M_{LL})_{12} & = & -s_{12} c_{12} c_{23} (m_1^{(1)} e^{-2i\beta_1}
- m_2^{(1)} e^{-2i\beta_2}) \rho \nonumber \\ & & - \left[ s_{12}
c_{12} c_{23} (m_1^{(2)} e^{-2i\beta_1} - m_2^{(2)} e^{-2i\beta_2}) -
m_3^{(0)} s_{23} k \, e^{i\delta} \right] \rho^2 + O(\rho^3) ,
\nonumber \\
(M_{LL})_{13} & = & s_{12} c_{12} s_{23} (m_1^{(1)} e^{-2i\beta_1} -
m_2^{(1)} e^{-2i\beta_2}) \rho \nonumber \\ & & + \left[ s_{12} c_{12}
s_{23} (m_1^{(2)} e^{-2i\beta_1} - m_2^{(2)} e^{-2i\beta_2}) +
m_3^{(0)} c_{23} k \, e^{i\delta} \right] \rho^2 + O(\rho^3) ,
\nonumber \\
(M_{LL})_{22} & = & s_{23}^2 m_3^{(0)} +
c_{23}^2 (m_1^{(1)}
s_{12}^2 \, e^{-2i\beta_1} + m_2^{(1)} c_{12}^2 \, e^{-2i\beta_2})
\rho \nonumber \\ & & + \,
c_{23}^2
(m_1^{(2)} s_{12}^2 \, e^{-2i\beta_1} + m_2^{(2)} c_{12}^2 \,
e^{-2i\beta_2})
\rho^2 +O(\rho^3) ,
\nonumber \\
(M_{LL})_{23} & = & s_{23} c_{23} \left[ m_3^{(0)} - (m_1^{(1)}
s_{12}^2 \, e^{-2i\beta_1} + m_2^{(1)} c_{12}^2 \, e^{-2i\beta_2}
) \rho \right. \nonumber \\ & & \left. \ \ \ - (m_1^{(2)}
s_{12}^2 \, e^{-2i\beta_1} + m_2^{(2)} c_{12}^2 \, e^{-2i\beta_2}
) \rho^2 \right] + O(\rho^3) ,
\nonumber \\
(M_{LL})_{33} & = & c_{23}^2 m_3^{(0)} +
s_{23}^2 (m_1^{(1)}
s_{12}^2 \, e^{-2i\beta_1} + m_2^{(1)} c_{12}^2 \, e^{-2i\beta_2} )
\rho \nonumber \\ & &
s_{23}^2
(m_1^{(2)} \! s_{12}^2 \, e^{-2i\beta_1} + m_2^{(2)} c_{12}^2 \,
e^{-2i\beta_2} )
\rho^2 + O(\rho^3) ,
\label{normelts}
\end{eqnarray} 
while the elements for the {\bf N2} and {\bf I} hierarchies read
\begin{eqnarray}
(M_{LL})_{11} & = & m_0 (c_{12}^2 \, e^{-2i\beta_1} + s_{12}^2 \,
e^{-2i\beta_2}) + m_2^{(2)} \! s_{12}^2 \, e^{-2i\beta_2} \rho^2 +
O(\rho^4) , \nonumber \\
(M_{LL})_{12} & = & -m_0 s_{12} c_{12} c_{23} (e^{-2i\beta_1} -
e^{-2i\beta_2}) \nonumber \\ & & + \! \left\{ m_2^{(2)} \! s_{12}
c_{12} c_{23} \, e^{-2i\beta_2} \! - \! k s_{23} \! \left[ m_0 \,
e^{-i\delta} (c_{12}^2 \, e^{-2i\beta_1} \! + \! s_{12}^2 \,
e^{-2i\beta_2}) - \! m_3 \, e^{i\delta} \right] \right\} \rho^2 \! +
O(\rho^4) , \nonumber \\
(M_{LL})_{13} & = & m_0 s_{12} c_{12} s_{23} (e^{-2i\beta_1} -
e^{-2i\beta_2}) \nonumber \\ & & - \! \left\{ m_2^{(2)} \! s_{12}
c_{12} s_{23} \, e^{-2i\beta_2} \! + \! k c_{23} \! \left[ m_0 \,
e^{-i\delta} (c_{12}^2 \, e^{-2i\beta_1} \! + \! s_{12}^2 \,
e^{-2i\beta_2}) - \! m_3 \, e^{i\delta} \right] \right\} \rho^2 \! +
O(\rho^4) , \nonumber \\
(M_{LL})_{22} & = & m_0 c_{23}^2 ( s_{12}^2 \, e^{-2i\beta_1} +
c_{12}^2 \, e^{-2i\beta_2} ) + m_3 s_{23}^2 \nonumber \\ & & + c_{12}
c_{23} \left[ m_2^{(2)} c_{12} c_{23} \, e^{-2i\beta_2} + 2m_0 s_{12}
s_{23} k \, e^{-i\delta} (e^{-2i\beta_1} - e^{-2i\beta_2}) \right]
\rho^2 + O(\rho^4) , \nonumber \\
(M_{LL})_{23} & = & -s_{23} c_{23} \left[ m_0 (s_{12}^2 \,
e^{-2i\beta_1} + c_{12}^2 \, e^{-2i\beta_2}) - m_3 \right] \nonumber
\\ & & -c_{12} \left[ m_2^{(2)} c_{12} s_{23} c_{23} \, e^{-2i\beta_2}
- m_0 s_{12} k \, e^{-i\delta} (c_{23}^2 - s_{23}^2)(e^{-2i\beta_1} -
e^{-2i\beta_2}) \right] \rho^2 + O(\rho^4) , \nonumber \\
(M_{LL})_{33} & = & m_0 s_{23}^2 (s_{12}^2 \, e^{-2i\beta_1} +
c_{12}^2 \, e^{-2i\beta_2}) + m_3 c_{23}^2 \nonumber \\ & & -c_{12}
s_{23} \left[ 2m_0 s_{12} c_{23} k \, e^{-i\delta} (e^{-2i\beta_1} -
e^{-2i\beta_2}) - m_2^{(2)} c_{12} s_{23} \, e^{-2i\beta_2} \right]
\rho^2 + O(\rho^4) . \label{invelts}
\end{eqnarray}
Several comments are in order at this point. First, note the
appearance of Majorana phases $\beta_{1,2}$ in the leading terms of
almost all $M_{LL}$ elements.  While it is tempting to neglect them
since they remain at present unrestricted by experiment, they enter in
a crucial way.  In particular, models frequently produce matrices with
$\det M_{LL}^0 \! < \! 0$, {\it i.e.}, some negative mass eigenvalues,
indicating a relative Majorana parity difference between neutrino
eigenstates.  Positive eigenvalues are rescued by adding $\pi/2$ to
the appropriate relative Majorana phases $\beta$, as can be seen from
Eqs.~(\ref{LHNM}), (\ref{MNSparam}), and
(\ref{MNSapprox})~\footnote{In contrast, similar sign differences
between eigenvalues of the Yukawa matrices are removed by redefining
Dirac fermion fields via $\psi \to \gamma_5 \psi$.}.

One popular texture~\cite{Haba} leading to bimaximal mixing within the
normal hierarchy resembles Eq.~(\ref{N1}), with the exception that
$(M_{LL})_{11} \! = \!  O(\rho^2)$.  From Eqs.~(\ref{normelts}) we see
that a very peculiar condition must occur in order for the $O(\rho)$
coefficient of $(M_{LL})_{11}$ to vanish: $(m_2^{(1)}/m_1^{(1)})
\tan^2 \theta_{12} \! = \! -e^{2i(\beta_2 - \beta_1)}$.  In fact,
typical textures with $(M_{LL})_{11} \! = \!  O(\rho^2)$ ({\it e.g.},
Ref.~\cite{ACM}) tend to lead to $\Delta m_{21}^2/|\Delta m_{32}^2| \!
= \! O(\rho^3)$, which are phenomenologically disfavored.  On the
other hand, the leading order of each element in the texture
Eq.~(\ref{N1}) remains unchanged even if the mass hierarchy is fully
hierarchical, {\it i.e.}, $m_1 \! = \!  O(\rho^2)$, or $m_1^{(1)} \!
= \! 0$.

The {\bf N2} and {\bf I} textures Eq.~(\ref{inv}), on the other hand,
reduce under certain circumstances to simpler forms.  If $(c_{12}^2 \,
e^{-2i\beta_1} + s_{12}^2 \, e^{-2i\beta_2}) \! = \! 0$, which is
accomplished by exactly ideal solar mixing ($\theta_{12} \! = \!
\pi/4$, which by Eq.~(\ref{solmix}) now appears disfavored by data;
note also Ref.~\cite{AFM}) and Majorana parities for the corresponding
mass eigenstates differing by $|\beta_1 \! - \! \beta_2| \! = \!
\pi/2$, then $(M_{LL})_{11} \! = \!  O(\rho^2)$.  Furthermore, if in
the {\bf I} texture one additionally has $m_3 \! = \! O(\rho)$ or
smaller, then the resulting texture assumes the form (corresponding to
Type IIA in Ref.~\cite{King}):
\begin{equation} \label{invsimp1}
M_{LL} \sim \left( \begin{array}{ccc}
\rho^2 & 1 & 1 \\ 1 & \rho^2 & \rho^2 \\ 1 & \rho^2 & \rho^2
\end{array} \right) \ .
\end{equation}
Unless $(c_{12}^2 \, e^{-2i\beta_1} + s_{12}^2 \, e^{-2i\beta_2}) \! =
\! 0$, however, the condition $m_3 \! = \! O(\rho)$ by itself does not
change the texture of $M_{LL}$ from the anarchic form Eq.~(\ref{inv}),
but it does make the lower $2\!\times\!2$ block singular at leading
order in $\rho$.  Alternately, if the first two mass eigenstates have
the same Majorana parity ($\beta_1 \!  = \! \beta_2$), then one
obtains the texture (corresponding to Type IIB in Ref.~\cite{King}):
\begin{equation} \label{invsimp2}
M_{LL} \sim \left( \begin{array}{ccc}
1 & \rho^2 & \rho^2 \\ \rho^2 & 1 & 1 \\ \rho^2 & 1 & 1
\end{array} \right) \ .
\end{equation}

Finally, the small value of $\theta_{13}$ [as indicated by the
appearance of $k$ in Eqs.~(\ref{normelts}) and (\ref{invelts})] leads
to its near insignificance in determining $M_{LL}$; it contributes to
no leading-order coefficients and almost disappears from the first
subleading terms as well.  In fact, the primary significance of the
smallness of $\theta_{13}$ is in its absence from a large number of
$M_{LL}$ elements, for which interesting model-independent relations
arise.  In particular, in the normal hierarchy [Eqs.~(\ref{normelts})]
the lower $2 \! \times \! 2$ block of $M_{LL}$ is singular at leading
order: $\left[ (M_{LL})_{22} (M_{LL})_{33} - (M_{LL})_{23}^2
\right]_{\rho=0} \! = \! 0$; but the same is true for the $O(\rho)$
coefficients of each of these elements by themselves, and similarly
for the $O(\rho^2)$ coefficients.  The last of these would be spoiled
if $U_{e3} \! = \! O(\rho)$ rather than $O(\rho^2)$.  A more
significant effect of $k$ arises in the (smaller) entries in the first
row and column of $M_{LL}$; for example, Eqs.~(\ref{normelts}) give
$(M_{LL})_{13}/(M_{LL})_{12} \! = \! -\tan \theta_{23}$---a direct
mixing observable---corrected only at relative $O(\rho)$ by $U_{e3}$.
If $U_{e3} \! = \! O(\rho)$, this relation is spoiled at leading
order.  The effects of $k$ in the inverted hierarchy are more
difficult to summarize in this way since they depend upon which
special case [{\it e.g.}, Eq.~(\ref{invsimp1}) or (\ref{invsimp2})] of
texture actually is used, but in no circumstance is $\theta_{13}$
revealed as the sole leading-order term of any $M_{LL}$ element.

From such considerations, we are led to the central result of this
work: A value of $\theta_{13} \! = \! O(\rho^2)$ or smaller, while
certainly possible phenomenologically, is difficult to obtain through
a model that predicts $M_{LL}$ directly, since such a small
$\theta_{13}$ does not contribute significantly to any element of
$M_{LL}$.  Its effect in $M_{LL}$ is masked by mixing with
$O(\rho^0)$, $O(\rho^1)$, and other $O(\rho^2)$ observables, and only
a model that predicts a very particular pattern of subleading effects
in the elements of $M_{LL}$ can naturally provide such a small
$\theta_{13}$.

\section{Normal Hierarchy} \label{NH}

All of our attention thus far has focused on the detailed texture of
$M_{LL}$.  Of course, in the seesaw mechanism $M_{LL}$ is not
fundamental but derived through Eq.~(\ref{seesaw}).  The question then
becomes whether it is possible to obtain textures for $M_{LL}$ as in
Eqs.~(\ref{normelts}) or (\ref{invelts}) through particular textures
of $M_{LR}$ and $M_{RR}$.  First, note that the seesaw formula
exhibits a symmetry: Inverting Eq.~(\ref{seesaw}) gives
\begin{equation}
M_{RR} = -M_{LR}^{\dagger} M_{LL}^{-1} M_{LR}^{\ast} \ .
\label{invseesaw}
\end{equation}
We see that this is the same as Eq.~(\ref{seesaw}) upon exchanging $LL
\! \leftrightarrow \! RR$ and $M_{LR} \! \leftrightarrow \!
M_{LR}^T$.  Let us now consider the extension of Point~(1) in
Sec.~\ref{conts} to the neutrino Yukawa mass matrix $M_{LR}$.  In
typical models, charged fermion Yukawa matrices include an $O(1)$ 33
element to represent that the Yukawa couplings leading to $m_t$,
$m_b$, and $m_\tau$ survive in the unbroken flavor symmetry limit.
This ansatz leads to phenomenologically successful small ratios for
$m_c/m_t$, $m_s/m_b$, and $m_\mu/m_\tau$; in the neutrino Dirac term,
however, there is no reason to anticipate such a large ratio (since
neutrino masses appear not directly from the Yukawa matrix but through
the seesaw mechanism).  Therefore we assume a texture for $M_{LR}$
that mirrors that of the charged fermions [compare Eq.~(\ref{ACL1}),
recalling that $\epsilon \! = \! O(\rho^2)$ and $\epsilon^\prime \! =
\! O(\rho^3)$], except that the 33 entry is no larger than the 22 or
23 entries:
\begin{equation} \label{nudirac}
M_{LR} \sim
\rho^2 \left ( 
\begin{array}{ccc}
l_{11}\rho & l_{12}\rho & l_{13}\rho \\
l_{21}\rho & l_{22} & l_{23} \\
l_{31}\rho & l_{32} & l_{33} 
\end{array} 
\right ) \ .
\end{equation}
We hasten to add that this ansatz for the magnitude of the 33 element
is not a crucial feature of this analysis, and discuss below the
effects of making the texture of $M_{LR}$ identical to that of the
charged fermions by setting $(M_{LR})_{33} \! = \! l_{33} \rho^0$
rather than $l_{33} \rho^2$.

Since $M_{LR}$ and $M_{LR}^T$ exhibit the same texture, one expects
similar textures for $M_{LL}$ and $M_{RR}$.  Interestingly, this is
not entirely true, as we now demonstrate.  Consider the determinant of
both sides of Eq.~(\ref{seesaw}); generically, $\det M_{LR}^\ast \! =
\!
\det M_{LR}^\dagger \! = \! O(\rho^7)$, while $\det M_{LL} \! = \!
O(\rho^2)$ from Eq.~(\ref{N1}), from which follows $\det M_{RR} \! =
\! O(\rho^{12})$.  In fact, the structure of $M_{RR}$ obtained directly
by combining $M_{LL}$ from Eq.~(\ref{N1}) and $M_{LR}$ from
Eq.~(\ref{nudirac}) yields leading-order terms
\begin{equation} \label{MRRderived}
M_{RR} \sim \rho^3 \left(
\begin{array}{ccc} a \rho^2 & bd \rho & cd \rho \\
bd \rho & b^2 \! + \! e \rho & bc \! + \! f \rho \\
cd \rho & bc \! + \! f \rho & c^2 \! + \! g \rho
\end{array} \right) \ ,
\end{equation}
where $a, \ldots , g$ are $O(1)$ coefficients that are complicated
functions of $l_{ij}$, the mixing angles, and the masses.  Of
particular note are that $(M_{RR})_{11}/(M_{RR})_{12} \! = \!
O(\rho)$, and that the last two columns (or rows) are proportional at
leading order, both in contrast to the texture of $M_{LL}$.  The
relative $O(\rho)$ corrections provided by $e , f, g$ lift the
leading-order singularity of $M_{RR}$, providing the required $\det
M_{RR} \! = \!  O(\rho^{12})$.  The overall $\rho^3$ factor in
$M_{RR}$ merely represents our ignorance of the scale of the largest
$M_{RR}$ element; the seesaw relation Eq.~(\ref{seesaw}) indicates
that each power of $\rho$ removed from $M_{RR}$ must return as a power
multiplying $M_{LL}$ in Eq.~(\ref{N1}).

It is important to reiterate that Eq.~(\ref{nudirac}), although
motivated by reasonable criteria, remains an ansatz.  If $M_{LR}$
possesses a number of elements smaller than suggested above (texture
zeroes) or a correlation among its leading $l_{ij}$ coefficients,
owing to the constraints of a flavor symmetry, then an $M_{RR}$
structure substantially different from Eq.~(\ref{MRRderived}) may be
mandated in order to obtain the desired $M_{LL}$.

In order to achieve the desired form of $M_{LL}$ with a generic
$M_{LR}$, a great deal of structure must be incorporated into
$M_{RR}$.  We demonstrate this by using {\em generic\/} $M_{LR}$ and
$M_{RR}$ textures of the forms given by Eqs.~(\ref{nudirac}) and
(\ref{MRRderived}) to compute $M_{LL}$.  The result of this exercise
does indeed give a texture of the form Eq.~(\ref{N1}), for which
$\left[ (M_{LL})_{22} (M_{LL})_{33} - (M_{LL})_{23}^2
\right]_{\rho=0} \! = \! 0$ [but not for the coefficients of the
$O(\rho)$ terms, as holds for Eqs.~(\ref{normelts})].  It also gives
mass eigenvalues in the normal hierarchy, $m_3 \! = \!  O(\rho^0)$,
$m_{1,2} \! = \!  O(\rho)$ and $\theta_{23} \! = \! O(\rho^0)$.
However, the value of $\theta_{13}$ obtained from these textures is
$O(\rho)$.  In order to obtain $\theta_{13} \! = \! O(\rho^2)$, it
would be necessary to modify Eq.~(\ref{MRRderived}) by the inclusion
of highly correlated subleading terms, which appears to violate the
spirit of naturalness and aversion to fine tuning in model building.

A few words are in order regarding the derivation of neutrino
observables mentioned in the previous paragraph.  It is not actually
necessary to solve the full eigenvector problem to determine the
leading contribution to most observables.  Rather, one begins by
reversing the argument above Eq.~(\ref{MRRderived}), to find first
that $\det M_{LL} \! = \!  O(\rho^2)$.  Next, since $M_{LL}$ is
approximately diagonalized by unitary $U_{\rm MNS}$ via
Eq.~(\ref{back}), then
\begin{equation}
{\rm Tr} M_{LL} \simeq {\rm Tr} ( U^\dagger_{\rm MNS} U_{\rm MNS}^\ast
M_{LL}^0) \ .
\end{equation}
Inasmuch as effects of the complex phase $\delta$ in $U_{\rm MNS}$ are
suppressed by the smallness of $s_{13}$, $U^\dagger_{\rm MNS} U_{\rm
MNS}^\ast$ is just the phase matrix $diag ( e^{-2i\beta_1},
e^{-2i\beta_2}, 1 )$, meaning that ${\rm Tr} M_{LL}$ is essentially
the sum of the eigenvalues, and ${\rm Tr} M_{LL}^2$ is essentially the
sum of their squares. $\det M_{LL}$, likewise, is the product of the
eigenvalues up to a phase.  These phases are inessential to our
discussion and hence are neglected.  Since in the present case ${\rm
Tr} M_{LL} \! = \! \mu \! + \! O(\rho)$, and the $O(\rho^0)$ term of
${\rm Tr} M_{LL}^2$ turns out to be $\mu^2$, one concludes that
precisely one eigenvalue is $O(\rho^0)$: $\mu \! = \! m_3^{(0)}
\Rightarrow m_3 \! = \! m_3^{(0)} + O(\rho)$, while $m_1$ and $m_2$ are
each $O(\rho^1)$.  The $O(\rho)$ coefficient $m_3^{(1)}$ can then be
determined since the $O(\rho)$ coefficient of ${\rm Tr} M_{LL}^2$ is
just $2m_3^{(0)} m_3^{(1)}$.

The leading-order terms of $m_3$ thus obtained may then be fed into
the equation $(M_{LL} \! - \! m_3 \openone) u_3 \! = \! 0$ to
determine the corresponding eigenvector $u_3$, which once normalized
is the third column of $U_{\rm MNS}$.  From this procedure one obtains
$(u_3)_2/(u_3)_3 \!  = \! O(\rho^0)$ and $(u_3)_1/(u_3)_3 \! = \!
O(\rho)$.  Normalization of $u_3$ and a glance at Eq.~(\ref{MNSparam})
then demonstrates that $\theta_{23} \! = \! O(\rho^0)$ and
$\theta_{13} \!  = \! O(\rho)$.

If, instead of the form Eq.~(\ref{nudirac}), one insists on a texture
for $M_{LR}$ completely parallel to that of the charged fermions [{\it
i.e.}, $(M_{LR})_{33} \! = \! l_{33} \rho^0$], then the powers of
$\rho$ in the texture for $M_{RR}$ change compared to
Eq.~(\ref{MRRderived}), but its rank and the basic predictions for
neutrino observables do not.  Specifically, one finds
\begin{equation}
M_{RR} \sim \frac 1 \rho \left(
\begin{array}{ccc} a \rho^6 & bd \rho^5 & cd \rho^3 \\
bd \rho^5 & (b^2 \! + \! e \rho) \rho^4 & (bc \! + \! f \rho) \rho^4 \\
cd \rho^3 & (bc \! + \! f \rho) \rho^2 & c^2 \! + \! g \rho
\end{array} \right) \ ,
\end{equation}
from which one still obtains $\theta_{23} \! = \! O(\rho^0)$ and
$\theta_{13} \! = \! O(\rho)$.  The overall $1/\rho$ indicates not a
singularity in the flavor limit, but again reveals the fact that the
inverse seesaw Eq.~(\ref{invseesaw}) allows the portability of overall
$\rho$ powers among $M_{LR}$, $M_{LL}$, and $M_{RR}$.

\section{Inverted Hierarchy} \label{IH}

For completeness, let us follow the same logic of using the inverted
seesaw mechanism [Eq.~(\ref{invseesaw})] for the {\bf N2} and {\bf I}
hierarchies, using the same ansatz [Eq.~(\ref{nudirac})] as above for
$M_{LR}$.  Starting with the explicit forms for the elements of
$M_{LL}$ given in Eqs.~(\ref{invelts}) (but with no further
constraints), one obtains the explicit form
\begin{equation}
M_{RR} \sim \rho^4 \left( \begin{array}{ccc} \rho^2 & \rho & \rho \\
\rho & 1 & 1 \\ \rho & 1 & 1 \end{array} \right) ,
\label{RRinv1}
\end{equation}
where comments from the previous section on the portability of the
overall factor of $\rho^4$ between $M_{LL}$ and $M_{RR}$ still
apply. One finds $\det M_{RR} \! = \! O(\rho^{14})$, the value
obtained through naive power counting on Eq.~(\ref{RRinv1}), as well
as the multiplicative nature of the determinant [$\det M_{LL} \! = \!
O(\rho^0)$, $\det M_{LR} \! = \! O(\rho^7)$] using
Eq.~(\ref{invseesaw}).  The lower $2\!\times\!2$ block is singular, it
turns out, only if $(c_{12}^2 \, e^{-2i\beta_1} + s_{12}^2 \,
e^{-2i\beta_2}) \! = \! 0$ [which leads to $M_{LL}$ as given in
Eq.~(\ref{invsimp1}), but again is disfavored by data], or if $l_{22}
l_{33} - l_{23} l_{32} \!  = \!  0$, {\it i.e.}, the lower
$2\!\times\!2$ block of $M_{LR}$ (which gives the leading-order
contribution to $\det M_{LR}$) is singular.  In particular, the
condition $\beta_1 \! = \! \beta_2$ by itself, which gives rise to the
texture Eq.~(\ref{invsimp2}), still gives a texture of the form
Eq.~(\ref{RRinv1}).

If $m_3 \! = \! O(\rho)$, then one obtains the form [{\it cf.}\
Eq.~(\ref{MRRderived})]
\begin{equation}
M_{RR} \sim \rho^3 \left( \begin{array}{ccc} d^2 \rho^2 & bd \rho & cd
\rho \\ bd \rho & b^2 & bc \\ cd \rho & bc & c^2 \end{array} \right) ,
\label{MRRinv}
\end{equation}
for which one finds $\det M_{RR} \! = \! O(\rho^{13})$ using
Eq.~(\ref{invseesaw}), owing to the extra power of $\rho$ in $\det
M_{LL}$ from $m_3$.  The naive power-counting result from
Eq.~(\ref{MRRinv}) is $\det M_{RR} \! = \! O(\rho^{11})$, but this
difference is resolved by the fact (apparent from this form) that
$M_{RR}$ is only rank 1.

If one changes Eq.~(\ref{nudirac}) to make $(M_{LR})_{33} = l_{33}
\rho^0$ in parallel to the charged fermion textures, then again the
specific powers emerging in the derived $M_{RR}$ change compared to
Eq.~(\ref{MRRinv}), but its rank and the size of the predicted
observables do not.  Specifically,
\begin{equation} \label{MRRinv2}
M_{RR} \sim \frac 1 \rho \left( \begin{array}{ccc} d^2 \rho^6 &
bd \rho^5 & cd \rho^3 \\ bd \rho^5 & b^2 \rho^4 & bc \rho^2 \\
cd \rho^3 & bc \rho^2 & c^2 \end{array} \right) .
\end{equation}

In each case, however, it is clear that the assumed value $s_{13} \! =
\! O(\rho^2)$ is buried even more deeply in the structure of $M_{RR}$
than in the normal hierarchy.  This is apparent already from
Eqs.~(\ref{invelts}), in which $k$ appears only in combination with
the second-order mass difference $m_2^{(2)}$.  The problem traces back
to larger [$O(\rho^0)$] values of the ``upper'' masses $m_{1,2}$ in
the inverted hierarchy, which tend to make the elements even in the
first row and column of $M_{LL}$ larger, thus obscuring the few places
where $s_{13}$ might hope to dominate.  Indeed, a calculation
analogous to that in the previous section shows that the mixing angle
$s_{13}$ obtained from either Eq.~(\ref{MRRinv}) or (\ref{MRRinv2}),
for example, is generically $O(\rho^0)$.
 
\section{Conclusions} \label{concl}

The low-energy neutrino observables, as determined from the latest
data, include two $O(1)$ mixing angles and a mass hierarchy ratio
which, in terms of the universal flavor symmetry-breaking parameter
$\rho \! = \!  O(0.1)$, is $O(\rho^2)$ ({\it i.e.}, not excessively
small).  This, we have seen, is sufficient to obscure a truly small
[$O(\rho^2)$] value for $|U_{e3}|$ emerging from a model that predicts
individual elements of the Majorana mass matrix $M_{LL}$.  In other
words, a certain amount of correlation between the elements of
$M_{LL}$, extending down to coefficients subleading in $\rho$, is
necessary in order to guarantee a generically small $|U_{e3}|$.  Such
correlations may be viewed as unappealing, a form of fine tuning.

We have illustrated the consequences of this scenario by computing the
explicit texture of $M_{LL}$ in both normal [Eqs.~(\ref{normelts})]
and inverted [Eqs.~(\ref{invelts})] hierarchies, using generic $O(1)$
values for solar and atmospheric angles and $O(\rho^2)$ for the
neutrino mass splittings ratio, and assuming a value of $|U_{e3}| \! =
\! O(\rho^2)$.  Although the most natural picture of flavor breaking
predicts an $O(\rho)$ contribution to $U_{e3}$ arising from
diagonalization of the charged lepton Yukawa matrix $Y_L$, we first
supposed that such a contribution is somehow suppressed.  We then
adopted a form for the neutrino Dirac mass matrix $M_{LR}$
[Eq.~(\ref{nudirac})] similar to those typically used for the charged
fermions.  Next, we employed the simple seesaw mechanism to derive
textures for the right-handed Majorana mass matrix $M_{RR}$, and asked
whether a texture of the {\em generic\/} form thus derived---without
excessive correlation of its subleading terms---naturally gives rise
to observables at the desired orders of magnitude.  In all cases, we
found that $O(\rho^2)$ values of $|U_{e3}|$ do not arise naturally, in
the sense of the criteria laid out in Sec.~\ref{conts}.

It is important to consider whether these criteria, or the form of the
Dirac matrix used in Eq.~(\ref{nudirac}), is too restrictive.  In
fact, it should be possible to evade these constraints if certain key
elements in $M_{LR}$ should turn out to be correlated at leading order
in $\rho$, or smaller than suggested by Eq.~(\ref{nudirac}).  An
appropriately restrictive flavor symmetry, for example, can provide
such useful correlations or texture zeroes.  One must nevertheless
keep in mind that these matrix elements are defined at high energy
scales at which the flavor symmetry breaks, and may mix through the
evolution of renormalization group equations down to low scales, thus
potentially obscuring such constraints.

Finally, we point out that the expressions derived here, with minor
modifications, can be used in another way.  Even though the purpose of
this paper has been the negative result of showing that $O(\rho^2)$
values of $|U_{e3}|$ are difficult to achieve, the results of
Eqs.~(\ref{normelts})--(\ref{invelts}) are still useful in the likely
physical situation, {\it i.e.}, $|U_{e3}| \! = \! O(\rho)$, if one
merely replaces each occurrence of $U_{e3} \! = \! k \rho^2
e^{-i\delta}$ by $k \rho \, e^{-i\delta}$.

\section*{Acknowledgments}
We thank E.~Geis for a reading of an early version of the manuscript,
and R.~Derm\'{i}\v{s}ek and S.F.~King for comments on the literature.
This work was supported by the National Science Foundation under Grant
No.\ PHY-0140362.

\end{document}